\documentclass{emulateapj}
\usepackage{graphicx}
\usepackage{txfonts}
\usepackage{natbib}
\usepackage[scaled]{helvet}
\usepackage{epsfig}
\usepackage{url}
\bibpunct{(}{)}{;}{a}{}{,}
\newcommand{\kms}{\,km\,s$^{-1}$}

\usepackage{color}
\usepackage[normalem]{ulem}

\begin{document}

\title{\bf An Extreme Analogue of $\epsilon$ Aurigae: An M-giant Eclipsed Every 69 Years by a Large Opaque Disk Surrounding a Small Hot Source}

\author{Joseph E. Rodriguez$^1$, Keivan G. Stassun$^{1,2}$, Michael B. Lund$^1$, Robert J. Siverd$^{3}$, Joshua Pepper$^{4}$, Sumin Tang$^{5,6,7}$, Stella Kafka$^{8}$, B. Scott Gaudi$^{9}$, Kyle E. Conroy$^1$, Thomas G. Beatty$^{10,11}$, Daniel J. Stevens$^{9}$, Benjamin J. Shappee$^{12,13}$, Christopher S. Kochanek$^{9,14}$}

\affil{$^1$Department of Physics and Astronomy, Vanderbilt University, 6301 Stevenson Center, Nashville, TN 37235, USA}
\affil{$^2$Department of Physics, Fisk University, 1000 17th Avenue North, Nashville, TN 37208, USA}
\affil{$^3$Las Cumbres Observatory Global Telescope Network, 6740 Cortona Dr., Suite 102, Santa Barbara, CA 93117, USA}
\affil{$^4$Department of Physics, Lehigh University, 16 Memorial Drive East, Bethlehem, PA 18015, USA}
\affil{$^5$Harvard-Smithsonian Center for Astrophysics, 60 Garden St, Cambridge, MA 02138, USA}
\affil{$^6$Division of Physics, Mathematics, and Astronomy, California Institute of Technology, Pasadena, CA 91125}
\affil{$^7$Kavli Institute for Theoretical Physics, University of California, Santa Barbara, CA 93106, USA}
\affil{$^{8}$American Association of Variable Star Observers, 49 Bay State Rd., Cambridge, MA 02138, USA}
\affil{$^{9}$Department of Astronomy, The Ohio State University, Columbus, OH 43210, USA}
\affil{$^{10}$Department of Astronomy \& Astrophysics, The Pennsylvania State University, 525 Davey Lab, University Park, PA 16802}
\affil{$^{11}$Center for Exoplanets and Habitable Worlds, The Pennsylvania State University, 525 Davey Lab, University Park, PA 16802}
\affil{$^{12}$Carnegie Observatories, 813 Santa Barbara Street, Pasadena, CA 91101, USA}
\affil{$^{13}$Hubble, Carnegie-Princeton Fellow}
\affil{$^{14}$Center for Cosmology and AstroParticle Physics (CCAPP), The Ohio State University, 191 W.\ Woodruff Ave., Columbus, OH 43210, USA}

\shorttitle{An Extreme Analogue of $\epsilon$ Aurigae}

\begin{abstract}
We present TYC 2505-672-1 as a newly discovered and remarkable eclipsing system comprising an M-type red giant that undergoes a $\sim$3.45 year long, near-total eclipse (depth of $\sim$4.5 mag) with a very long period of $\sim$69.1 yr. TYC 2505-672-1 is now the longest-period eclipsing binary system yet discovered, more than twice as long as that of the currently longest-period system, $\epsilon$ Aurigae. We show from analysis of the light curve including both our own data and historical data spanning more than 120 yr and from modeling of the spectral energy distribution, both before and during eclipse, that the red giant primary is orbited by a moderately hot source ($T_{\rm eff} \approx 8000$ K) that is itself surrounded by an extended, opaque circumstellar disk. From the measured ratio of luminosities, the radius of the hot companion must be in the range 0.1--0.5 R$_\odot$ (depending on the assumed radius of the red giant primary), which is an order of magnitude smaller than that for a main sequence A star and 1--2 orders of magnitude larger than that for a white dwarf. The companion is therefore most likely a ``stripped red giant" subdwarf-B type star destined to become a He white dwarf. It is however somewhat cooler than most sdB stars, implying a very low mass for this ``pre-He-WD" star. The opaque disk surrounding this hot source may be a remnant of the stripping of its former hydrogen envelope. However, it is puzzling how this object became stripped, given that it is at present so distant (orbital semi-major axis of $\sim$24 AU) from the current red giant primary star. Extrapolating from our calculated ephemeris, the next eclipse should begin in early UT 2080 April and end in mid UT 2083 September (eclipse center UT 2081 December 24). In the meantime, radial velocity observations would establish the masses of the components, and high-cadence UV observations could potentially reveal oscillations of the hot companion that would further constrain its evolutionary status. In any case, this system is poised to become an exemplar of a very rare class of systems, even more extreme in several respects than the well studied archetype $\epsilon$ Aurigae.
\end{abstract}


\section{Introduction}
One of the most well studied eclipsing binaries (EB) is  $\epsilon$~Aurigae (HD 31964). At $V$ $\sim$ 3 and having the longest known orbital period for an EB ($\sim$27.1 yr), this unique system has become a prime target for extensive characterization. The primary eclipse has a depth of 0.8--1.0 mag (visual) and lasts for $\sim$2 yr. The primary star is an evolved F0 giant first proposed as being eclipsed by a very large faint companion \cite{Carroll:1991}. The Spectral Energy Distribution (SED) of $\epsilon$ Aur was reproduced using 2 components: a 2.2 M$_\sun$ post-asymptotic giant branch F star, and a 5.9 M$_\sun$ B5V star with a thick semi-transparent disk \citep{Hoard:2010}. Using the CHARA array to obtain interferometric images during the 2009-2011 eclipse, \citet{Kloppenborg:2010} confirmed the eclipse to be caused by a dark companion with a tilted disk.

In this work, we present the analysis of TYC 2505-672-1, a system similar to $\epsilon$ Aur, but with an even longer period of $\sim$69.1 yr, making it now the EB with the longest known period. We use catalog photometry fortuitously obtained both during and prior to eclipse for an analysis of the system spectral energy distribution (SED), and we use extensive photometric observations from the Kilodegree Extremely Little Telescope (KELT) together with archival observations spanning 120 yr. The primary component of the system is an M-type red giant that over the past century has shown two very deep, multi-year-long dimming events, most recently noted in Astronomer Telegrams by the MASTER Global Robotic Net \citep{Lipunov:2010}. It has been suggested that the dimmings are caused by either R Coronae Borealis (RCB) events of the M-giant \citep{Denisenko:2013} or by a very long-period eclipse of the M-giant by a large, faint companion as in $\epsilon$ Aur \citep{Tang:2013}. 

From our SED and light curve analysis, we interpret the dimmings to be caused by a small, hot companion surrounded by a large opaque disk eclipsing the M-giant primary star every $\sim$69 yr. However, as we discuss, the evolutionary status of this hot companion is unclear, but may be a rare example of a low-mass, recently ``stripped red giant" destined to become a Helium white dwarf, such as that reported by \citet{Maxted:2014}.

\section{Characteristics of the TYC 2505-672-1 System}
The known properties of the TYC 2505-672-1 (2MASS J09531000+3353527) system ($\alpha$ = 09h 53m 10.0043s, $\delta$ = $+33^{\circ}$ 53$\arcmin$ 52.734$\arcsec$; V$\sim$10.71) are a bit sparse \citep{Hog:1998, Hog:2000}. \citet{Afanasiev:2013} observed the optical spectra of TYC 2505-672-1 during the dim state and found it to be consistent with an M1 III red giant. They did observe H-alpha emission in the spectra and suggest that the M-giant might be entering an RCB phase. \citet{Pickles:2010} found from spectral template fitting a best-fit spectral type of M2 III; in order to be as conservative as possible in estimating the stellar and system parameters, we adopt a very broad range of spectral types (M0-8IIIe) for the primary star in the analysis that follows.

\section{Data}
Over the past century, multiple surveys have observed TYC 2505-672-1 at a variety of cadences (see Figure \ref{figure:FullLC}). Note that over the $\sim$120 yr time span of the data there have been two apparent eclipses, one recently in 2011--2015, and one sparsely sampled around 1942--1945. We next describe these photometric light curves, and the available catalog broadband absolute photometric data, in turn.

\subsection{KELT-North}
The Kilodegree Extremely Little Telescope (KELT-North) is an ongoing photometric survey searching for transiting planets around bright ($V$ = 8-11)  stars. KELT-North uses a Mamiya 645-series wide-angle lens with a 42mm aperture and a 80mm focal length (f/1.9), corresponding to a large field of view ($26^{\circ}$ $\times$ $26^{\circ}$) with a plate scale of 23$\arcsec$ per pixel. The telescope has a non-standard filter, comparable to an extra-broad R-band, with a typical photometric RMS precision of $\textless$1$\%$ for bright stars, but varies substantially across the KELT field. The survey observes a predefined set of fields with a $\sim$15 minute cadence through the entire season of visibility of each field \citep{Pepper:2007}. TYC 2505-672-1 is located in KELT-North Field 06, which is centered on ($\alpha$ =  09hr 46m 33.752s, $\delta$ = $+31^{\circ}$ 39$\arcmin$ 24.11$\arcsec$). KELT-North observed this field from UT 2006 October 27 to UT 2014 December 21, obtaining 9,320 images. The data were reduced using a heavily modified version of the ISIS software package, described further in \S2 of \citet{Siverd:2012}. The photometric scatter (outside the eclipse) of the KELT-North light curve for TYC 2505-672-1 is $\sim$2$\%$, roughly consistent with the expected scatter for a target of this brightness located at its position in the KELT-North field. Observations during the eclipse are at the observational limit of KELT-North. Therefore, we do not trust the observed in-eclipse variability from the KELT-North data. 

\subsection{American Association of Variable Star Observers (AAVSO)}
The Association of Variable Star Observers (AAVSO) is a worldwide network of amateur and professional astronomers dedicated to the understanding of variable stars. AAVSO monitored TYC 2505-672-1 from UT 2013 February 08 until UT 2015 September 22, obtaining 246 observations in $V$ band (and visual observations). The observations presented in this work were taken by 18 different observers from the AAVSO network. Many of the AAVSO members use an web interface photometry tool on the AAVSO website called Variable star PHOtometry Tools (VPHOT). The average error from all observers is 0.02 mag with a standard deviation of 0.35 mag.

\subsection{Digital Access to a Sky Century at Harvard (DASCH)}
The Digital Access to a Sky Century at Harvard (DASCH) survey is a digitized version of the Harvard astronomical photographic plate collection. These observations allow the astronomical study of objects on the century-long time scale. To date, they have scanned over 100,000 plates corresponding to over 7 billion measured magnitudes. The DASCH observations are in the $B$ bandpass and have limiting magnitude of 15 (this value does vary). The DASCH data release 4 represents observations from 1885 to 1992 (see \citet{Grindley:2012} for an overview of the survey). The DASCH survey observed TYC 2505-672-1 from UT 1890 March 08 until UT 1989 December 01, obtaining 1432 observations. Only some of the observations have listed errors. The average of the listed errors is 0.1 mag with a standard deviation of 0.03 mag.

\subsection{Catalina Real-time Transient Survey (CRTS)}

The Catalina Real-time Transient Survey (CRTS) is a wide photometric survey consisting of 3 telescopes covering 33,000 Deg$^{2}$ to find rare transient objects. All transient objects are openly published within minutes of the observations. See \citet{Drake:2009} for imformation about the survey and data reduction process. CRTS observed TYC 2505-672-1 from UT 2006 February 22 until 2013 June 05, resulting in 78 measurements. The photometric values are determined using the SExtractor software package \citep{Bertin:1996}. The average error for the CRTS observations is 0.055 mag with a standard deviation of 0.005 mag.

\subsection{All-Sky Automated Survey for SuperNovae (ASAS-SN)}
The All-Sky Automated Survey for SuperNovae (ASAS-SN or ``Assassin''; \citet{Shappee:2014}) is a long-term project to monitor the whole sky down to a limiting magnitude of $V \sim 17$ mag with the highest cadence possible using a global network of telescopes with a modular design.  The focus of the survey is to find nearby supernovae (SNe) and other bright transient sources.  Currently, ASAS-SN consists of two fully robotic units on Mount Haleakala in Hawaii and Cerro Tololo in Chile.  Each unit has four telescopes on a common mount and is hosted by Las Cumbres Observatory Global Telescope Network.  Each telescope consists of a 14-cm aperture Nikon telephoto lens and a 2k $\times$ 2k thinned CCD, giving a $4.5\times4.5$ degree field-of-view and a 7$\farcs$8 pixel scale.  These 8 telescopes allow ASAS-SN to survey 20,000~deg$^2$ per night, covering the entire visible sky every two days.  The pipeline is fully automatic and discoveries are announced within hours of the data being taken.  ASAS-SN has observed the field containing TYC~2505-672-1 141 times since UT 2012 January 23.  For the ASAS-SN data, we remove epochs affected by clouds and performed aperture photometry using the IRAF {\tt apphot} package and calibrated the results using the AAVSO Photometric All-Sky Survey (APASS; \citealp{Henden:2015}). The average ASAS-SN error for TYC~2505-672-1 is 0.022 mag with a standard deviation of 0.021 mag. 

\begin{figure*}[!ht]
\centering\epsfig{file=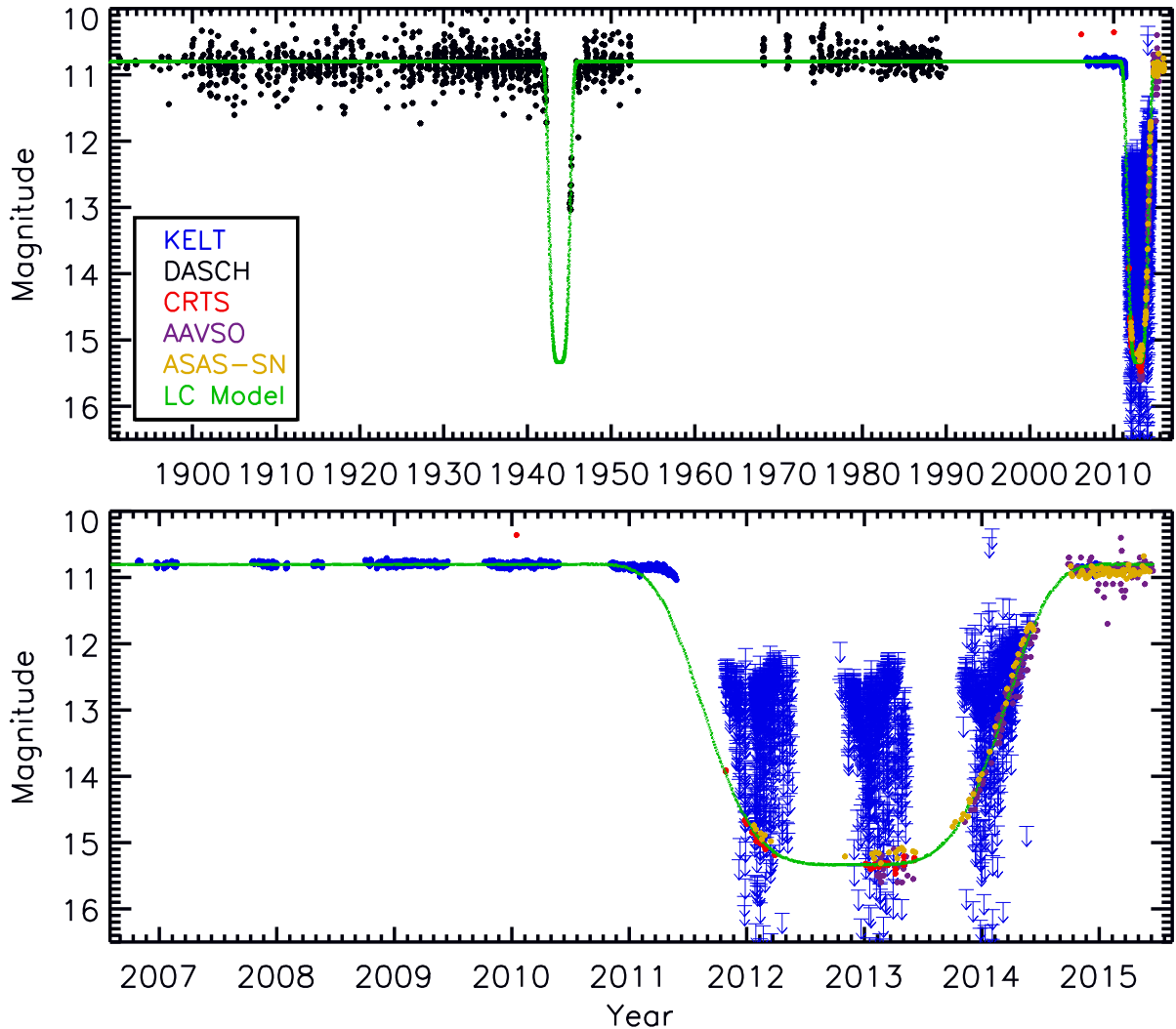,clip=,width=0.9\linewidth}
\caption{(Top) The KELT-North (Blue), DASCH (Black), CRTS (Red), AAVSO (Violet), and ASAS-SN (Yellow) observations plotted from 1890 to 2015. The green line represents a LC model of the combined photometric data. (Bottom) The photometric observations covering the most recent eclipse. The KELT-North observations during the eclipse are below the faintness limit of KELT and are therefore only upper limits. Only the AAVSO, CRTS, and ASAS-SN data are in the Visual and V-band magnitudes. We approximate the all observations to the AAVSO V-band to match the quiescent magnitude of the AAVSO data but no attempt has been made to place all the data on the same absolute scale. }
\label{figure:FullLC}
\end{figure*}
\subsection{Broadband Photometry from the Literature for Spectral Energy Distribution Modeling}

In order to ascertain the physical nature of the system, and in particular to help constrain the properties of the occulting body, we assembled all of the available photometry from the literature, which we then use in Section \ref{sec:sed} to model the spectral energy distribution (SED) of the system. All the broadband measurements are listed in Table \ref{tbl:SED}, and they are organized for convenience according to whether the available measurements happened to be obtained during occultation or not.

\begin{table}[ht]
\centering
\caption{Archival flux measurements of TYC 2505-672-1 used in the SED analysis.\label{tab:seddata}}
\begin{tabular}{ | c | c | c | c | c |}
\hline
Band & Magnitude & Error\tablenotemark{a} & Source & Reference\\
\hline
FUV  &  21.07   &  0.29  & GALEX & \citet{Bianchi:2011}\\
NUV    & 19.476   & 0.1& GALEX & \citet{Bianchi:2011}\\
$u'$   & 14.778   & 0.05  & SDSS &  \citet{Pickles:2010}\\
$g'$   & 11.501    &0.05 & SDSS & \citet{Pickles:2010}\\
$r'$   &  10.181    & 0.05 & SDSS & \citet{Pickles:2010}\\
$z'$   & 9.575   & 0.05 & SDSS & \citet{Pickles:2010}\\
BT     & 13.128   & 0.279  & Tycho-2 & \citet{Hog:2000} \\
VT     & 10.938   & 0.061  & Tycho-2 & \citet{Hog:2000}\\
$J$    &  7.614   & 0.05   & 2MASS & \citet{Cutri:2003}\\
$H$     & 6.781   & 0.05 & 2MASS & \citet{Cutri:2003}\\
$K$    &  6.567   & 0.05 & 2MASS & \citet{Cutri:2003}\\
WISE1   & 9.179   & 0.065 & WISE & \citet{Cutri:2014}\\
WISE2  &  9.859   & 0.05 & WISE&  \citet{Cutri:2014}\\
WISE3  & 11.559   & 0.1 & WISE & \citet{Cutri:2014}\\
WISE4  & 12.386  &  0.05 & WISE & \citet{Cutri:2014}\\
\hline
In-Eclipse & & & &\\
\hline
$B$     &  16.382  &  0.05   & APASS & \citet{Henden:2015}\\
$V$    &  15.032  &  0.052 & APASS&  \citet{Henden:2015}\\
$g'$ &  15.711  &  0.05 & APASS&  \citet{Henden:2015}\\
$r'$ &  14.544  &  0.197 & APASS&  \citet{Henden:2015}\\
$i'$ &  13.755 &   0.201 & APASS&  \citet{Henden:2015}\\
\hline
\end{tabular}
\label{tbl:SED}
\begin{flushleft}
  \footnotesize \textbf{\textsc{NOTES}} \\
  \footnotesize $^a$Single-epoch errors have been inflated to reflect time variability of the source.
  \end{flushleft}
\end{table}
\section{Results}

\subsection{SED Analysis and Implications}
\label{sec:sed}
As shown in Table \ref{tbl:SED}, we are fortunate to have broadband photometry from the literature both outside of occultation and during occultation, at wavelengths from the GALEX FUV band (0.15$\mu$m) to the WISE4 band (20$\mu$m), providing a rich dataset for modeling the underlying component(s) of the system. As we discuss in Section \ref{sec:interp}, our modeling  of the SED conclusively shows that there is a small hot star in the system (possibly a white dwarf), and that this small hot star is likely to be surrounded by a large cool disk. 

\begin{figure*}[!ht]
\centering\epsfig{file=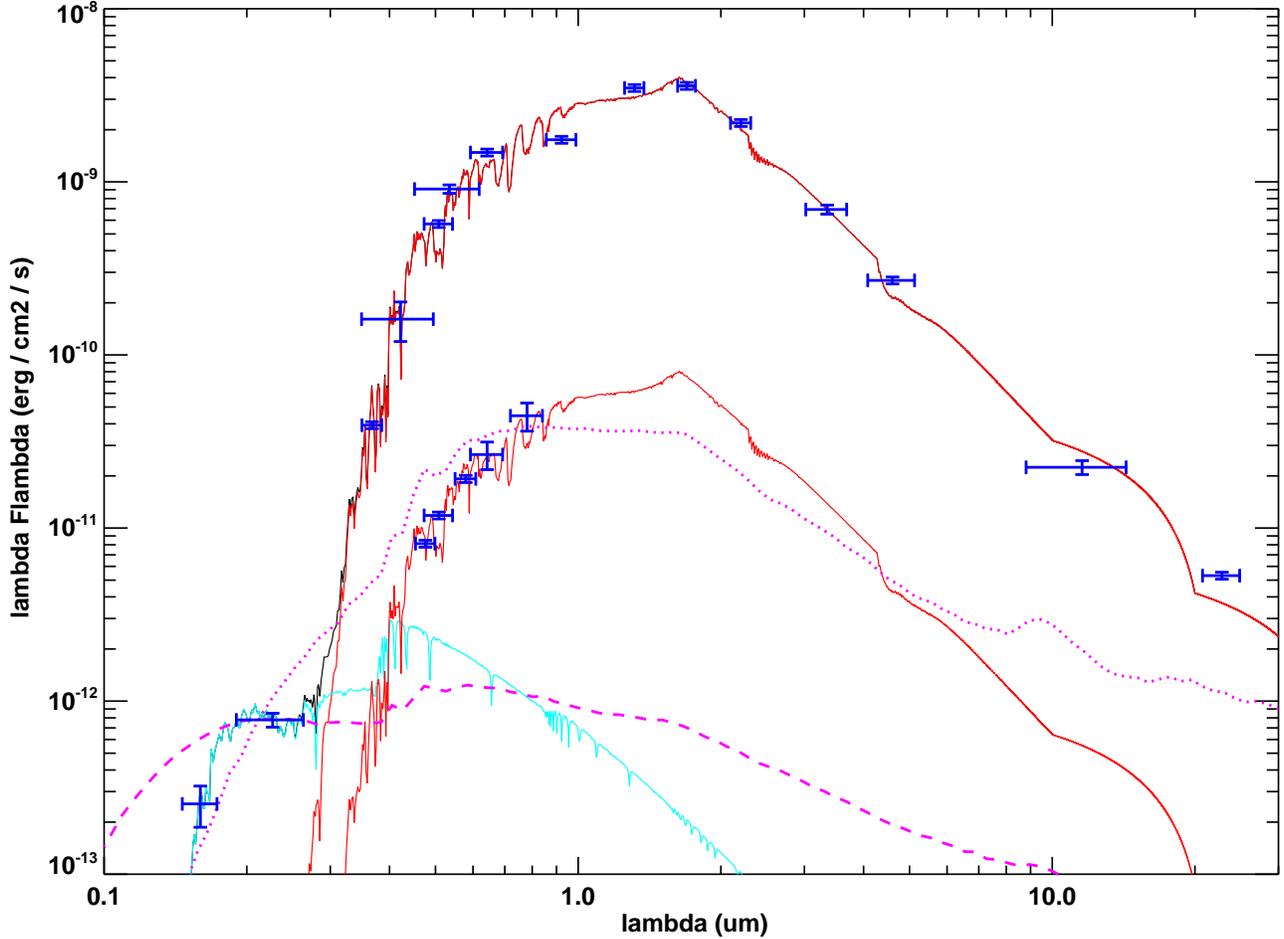,angle=90,clip=,width=0.99\linewidth}
\caption{Spectral Energy Distribution fit for TYC 2505-672-1. The upper red curve has $T_{\rm eff} = 3600$ K while lower red curve is the same model but scaled down by a factor of 50. The blue curve is the best fit to the GALEX fluxes; it has $T_{\rm eff} = 8000$ K. The magenta dashed curve is a low-mass M-dwarf with an accretion rate of 10$^{-6}$ M$_\odot$ yr$^{-1}$. The dotted magenta curve shows what it would take for a cool star with a low accretion rate to match the GALEX points (a solar-type star accreting at 3 $\times$ 10$^{-8}$ M$_{\odot}$ yr$^{-1}$). }
\label{SED_Fit}
\end{figure*}

We fit three separate Kurucz atmosphere models to the available data. First, we fit a cool, low gravity model ($\log g = 2.5$, as appropriate for a modestly-evolved red giant) to the data obtained outside of eclipse, excepting the GALEX fluxes. Second, we fit the same model to the data obtained during eclipse. Third, we fit a hot source to the GALEX fluxes, with the additional constraint that the sum of this hot source and that of the first step are consistent with the SDSS $u$-band measurement. In each model fit, the fit parameters were the effective temperature, the extinction, and a normalization. Note that according to the Galactic dust maps of \citet{Schlegel:1998}, the maximum extinction for this line of sight is $A_V = 0.04$ mag, therefore the precise extinction value is of minor importance. We adopted solar metallicity for simplicity; these broadband fits are not strongly sensitive to the choice of metallicity.

The resulting best SED fits are shown in Figure \ref{SED_Fit}. The upper red curve has $T_{\rm eff} = 3600$ K as appropriate for a red giant, and consistent with the spectral class of M2 III found by \citet{Pickles:2010}. The lower red curve is the same model but scaled down by a factor of 50. The blue curve is the best fit to the GALEX fluxes and to the $u$-band flux; it has $T_{\rm eff} = 8000$ K, such as for a cool white dwarf. 

It is possible that the small excess apparent in the SED at 20 $\mu$m is due to thermal infrared emission from the disk around the companion star. It is beyond the scope of this paper to model such a disk, given the lack of observational constraints on the disk emission. However, if the disk emits strongly as a nearly ``flat-spectrum" source then its emission at 20 $\mu$m would be on the order of $\sim$10$^{-12}$ erg cm$^{-2}$ s$^{-1}$ (based on the peak emission of the companion), which at 20 $\mu$m is $\sim$20\% of the red giant's photospheric emission and thus could plausibly account for the modest excess emission observed at that wavelength. Observations in the near- to mid-IR during eclipse of the red giant primary would definitively test this possibility. 

Our SED analysis provides the following results and interpretations:

(1) UV fluxes. The fact that the system is detected in both the GALEX NUV and FUV bands clearly indicates the presence of a hot component in the system; an M star alone cannot explain this UV excess emission. As can be seen from the SED fit, a secondary star with $T_{\rm eff} = 8000$ K fits the two GALEX fluxes nicely. It is possible that the UV flux is arising from something other than a stellar photosphere. Specifically, if accretion is occurring in the disk around the companion, this could cause a UV flux from photons inside the disk being scattered and escaping. If the observed UV flux is from accretion onto a cool star, then the photospheric emission of the star will be lower than that of the hot component shown in our SED fit, which would then require a very high accretion rate to reproduce the observed UV flux. Utilizing the SED models of low mass stars with accreting disks from \citet{Robitaille:2006}, we attempted to fit the GALEX fluxes with a low-mass star that is actively accreting from a disk. While a comprehensive search of all possible parameters is beyond the scope of this paper, in general we found that it is not possible to simultaneously fit both GALEX fluxes with such a model and a reasonably low accretion rate. A stellar photosphere with a low accretion rate fitting the GALEX fluxes would require the peak of the photospheric SED to rise far above the blue curve in Figure \ref{SED_Fit} (the magenta dotted line represents a solar-type star accreting at 3 $\times$ 10$^{-8}$ M$_{\odot}$ yr$^{-1}$), which would then be inconsistent with the observed SED in eclipse. In order to keep the peak of the photospheric SED low, then the shape of the SED must be relatively flat, such as that shown by the magenta dashed curve, which requires an M-dwarf with a high accretion rate of 10$^{-6}$ M$_\odot$ yr$^{-1}$, which would then deplete the disk on a very short timescale. 


(2) Recent and historical eclipses must both be the primary eclipse; there is not yet an observed secondary eclipse. If the historical eclipse were interpreted as the secondary eclipse, then one could infer the ratio of $T_{\rm eff}$ from ratio of the eclipse depths. From the full observed light curve (Figure \ref{figure:FullLC}), we may hypothesize that we are seeing two eclipses, a primary eclipse with a depth of $\sim$5 mag that has just recently occurred, and a secondary eclipse with a depth of $\sim$2 mag that occurred 70 years ago. The durations of the two eclipses are similar (about 4 years long), which would suggest a nearly circular orbit. In that case, the ratio of eclipse depths (in flux units) is approximately the ratio of surface brightnesses of the two bodies. We would have in this case a ratio of 100:6, which would imply a $T_{\rm eff}$ ratio of $\sim 15^{1/4} \approx 2$. This is in fact quite close to the ratio of $T_{\rm eff}$ from the SED fitting above ($8000 / 3600 \approx 2$). 

Another constraint is the ratio of luminosities from the primary eclipse depth. Assuming again that both a primary and secondary eclipse are observed, and that the primary eclipse is near total (which it appears to be from the roughly flat bottom), then the primary eclipse would represent a total blocking of the smaller body by the larger one. The ratio of light lost to light remaining at the bottom of the eclipse is then the ratio of luminosities of the two bodies. In this case, with a primary eclipse depth of $\sim$4.5 mag, we have a luminosity ratio of $\sim$100. 

\begin{figure*}[!ht]
\centering\epsfig{file=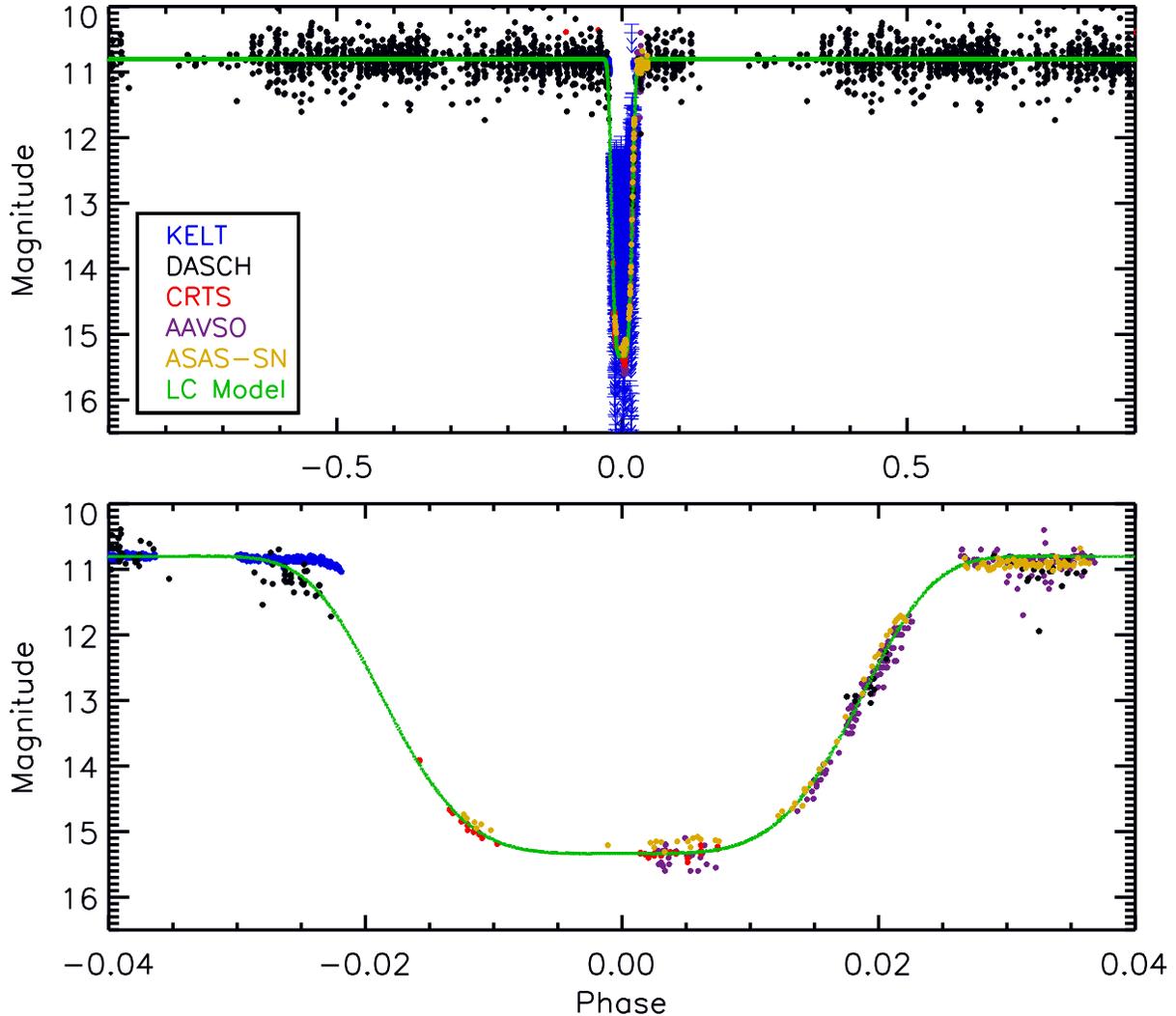,clip=,width=0.9\linewidth}
\caption{(Top) KELT-North (Blue), DASCH (Black), CRTS (Red), AAVSO (Violet), and ASAS-SN (Yellow) lightcurves phased to a period of 69.068 years (Bottom) Zoom in of the eclipse. The green line represents a LC model of the combined photometric data. The KELT-North observations during the eclipse are below the faintness limit of KELT and are therefore only upper limits. For a better visual representation of the in-eclipse structure, the KELT upper limit observations are not included in the bottom figure. Only the AAVSO, CRTS, and ASAS-SN data are in the Visual and V-band magnitudes. We approximate the all observations to the AAVSO V-band to match the quiescent magnitude of the AAVSO data but no attempt has been made to place all the data on the same absolute scale.  }
\label{figure:phased}
\end{figure*}

In order for all of the above to be internally consistent, the fully eclipsed body would have to be both the hotter object {\it and} the more luminous one. However, as can be seen from the SED (Figure \ref{SED_Fit}), the hot component (blue curve) is only more luminous than the red one at UV wavelengths. At visible wavelengths the red giant component dominates by a very large factor. Instead, the observed GALEX fluxes must represent the unobstructed fluxes of the hot component, because it would have to be fully blocked behind the red giant if it is the eclipsed body at primary eclipse. However, in that case GALEX would not have detected the hot component. This then severely limits how luminous the hot component can be relative to the red giant, and implies that it is the red giant that is eclipsed at primary eclipse, and that the data do not show any evidence of a visible secondary eclipse. Indeed, the recently observed eclipse and the historically observed eclipse both phase together nicely (Fig.\ \ref{figure:phased}), consistent with them representing the same primary eclipse separated by $\sim$69.068 yr.

A possible solution is that the hot component is surrounded by a large, cool disk, and that this is the body that obscures the red giant at primary eclipse. In that case, the red giant would simply become much fainter during eclipse (corresponding to the red curve in Figure \ref{SED_Fit} that matches the APASS SED during eclipse), as a result of being blocked by a large occulting screen. Indeed, in the faint state, the SED appears dominated by the same red giant spectrum as in the bright state, only diminished by a factor of $\sim$50, consistent with the same dominant light source being mostly blocked by a dark screen. Moreover, the occulting screen evidently produces a nearly grey extinction, since the shape of the SED of the red giant component in the faint state is not reddened. We estimate the physical dimensions of the disk surrounding the hot component in Section \ref{sec:Finterp}.


Finally, we can measure the ratio of the stellar radii from the Stefan-Boltzmann law, using the measured ratio of luminosities from the SED fits and the ratio of the best-fit temperatures: $R_{\rm hot} / R_{\rm RG} = \left[ \left(F_{\rm bol,hot} / F_{\rm bol,RG}\right) / \left(T_{\rm eff,hot} / T_{\rm eff,RG}\right)^4 \right]^{1/2}$, where the ``hot" and ``RG" subscripts refer to the hot companion and the red giant primary, respectively. The ratio of bolometric fluxes, $F_{\rm bol}$ is obtained simply by integrating the best-fit SEDs over all wavelengths, namely 0.00022. The resulting radius ratio is $\approx$0.003. Assuming a radius for the red giant primary in the range of 45--170 R$_\odot$ (depending on the assumed mass and age of the red giant; see below), this translates into a range of radii for the hot companion of 0.13--0.51 R$_\odot$. This calculation assumes a thermal, photospheric source and that we are seeing its entire surface (secondary star). It is possible that part of the secondary star is obscured by the disk around it causing us to underestimate its radius. The uncertainty may be as large as a factor of $\sim$2, given the uncertain $T_{\rm eff}$ from the SED fitting. This estimate for the radius of the hot companion is an order of magnitude smaller than the expected radius for a main-sequence A type star ($T_{\rm eff} \sim 8000$ K, $R \sim 2$ R$_\odot$), and 1--2 orders of magnitude larger than that expected for standard, cooling white dwarfs ($\approx$0.003--0.03 R$_\odot$, depending on mass). 

If the companion is actually a cooler star with accretion, then the temperature of the companion is lower than we have estimated here and consequently its radius would be larger, perhaps consistent with a standard main sequence cool dwarf. However we do not consider this likely because of the high accretion rate it would require (see result (1) earlier in this section). The in-eclipse SED (both optical and UV) can also be fit using a solar photosphere with a low accretion value of 3 $\times$ 10$^{-8}$ M$_\odot$ yr$^{-1}$ (magenta dotted curve in Figure \ref{SED_Fit}). In this scenario the companion radius would be $\sim$1 R$_\odot$. In other words, the dotted magenta curve suggests that another possible interpretation of the SED during eclipse is that the red giant primary is 100\% extinguished by the disk and that the companion SED is that of a solar-type star accreting at 3 $\times$ 10$^{-8}$ M$_{\odot}$ yr$^{-1}$ (the magenta dotted curve in Figure \ref{SED_Fit}). However, this model does not fit the in-eclipse SED as well as our preferred model, in which the red giant remains partially visible during eclipse (the lower red curve in Figure \ref{SED_Fit}) and the UV flux is provided by a small hot source, which fits the UV part of the SED extremely well. We discuss below the likelihood that the hot component is instead a ``stripped red giant" sdB type star.

\subsection{Orbital Period}
From Figure \ref{figure:FullLC} and the SED analysis in Section~\ref{sec:sed}, we interpret the observations of the two eclipses to be the primary eclipse observed twice. Given the depth of the recent event ($\sim$4.5 mag), it is possible that TYC 2505-672-1 dimmed below the limiting magnitude of the DASCH plates ($B$ $\sim$15 mag). We can also rule out the possibility that the eclipse happens every $\sim$34.5 years since we would have seen two additional events around $\sim$1908 and $\sim$1979, where we have sufficient coverage to rule out eclipses.

In order to calculate the period, we used a generalized normal distribution to find the midpoint of the event. A generalized normal distribution provides a good functional fit to a transit event without relying on any physical models, and the only physical parameters that are directly measured are the out-of-transit magnitude and the midpoint of transit. For the more recent event, we combined the light curves from AAVSO and CRTS into a single light curve, and then used a least-squares-fit optimization to fit a generalized normal distribution to this data to find that the midpoint of the event is 2456261.12224$\pm$2.081 days. We then fit the same function to the DASCH data, allowing only the baseline magnitude and midpoint of transit to be changed and preserving the shape of the transit. For the older event, we found a midpoint of the event of 2431033.91053$\pm$4.862 days. Using these two event midpoints, we calculate the event as having a period of 69.068$\pm$0.019 years. The initial dimming observed by KELT in mid 2011 does not line up with this symmetric eclipse model fully which might indicate that the eclipse eclipse is not symmetric in shape. 

\section{Interpretation and Discussion}
\label{sec:interp}


\subsection{Favored Interpretation: A Red Giant Eclipsed by a Pre--Helium-White-Dwarf Companion Surrounded by a Large Opaque Disk}
\label{sec:Finterp}
From the SED analysis, we have determined that this system is composed of an M-giant primary star with a hot (T$_{eff}\sim$8000 K) companion that is not contributing a significant amount of optical light. This secondary component could be a main-sequence A-type star or a cool white dwarf.  However, as discussed above, the apparent radius of the hot component is much too large to be a standard  dwarf and much too small to be a main-sequence star. 

Subdwarf B (sdB) stars are usually interpreted as red giants that been stripped of their hydrogen envelopes, leaving behind an exposed, hot core with an O or B type temperature but with a smaller radius than that of a main-sequence O or B type dwarf and a much larger radius than that of a hot white dwarf. It is expected that these objects eventually become Helium  dwarfs (i.e., they are ``pre-He-WD"). In this case, however, the temperature  is unusually cool for this scenario even if the radius is consistent. We note that \citet{Maxted:2014} reported a pre-He-WD system with a cool temperature also of $\sim$8000 K.

Therefore, we suggest that the most plausible interpretation of this system is an eclipsing binary with an M-giant primary and pre-He-WD companion that is surrounded by a large disk. This scenario explains the observed UV excess, the small contribution of the companion to the optical fluxes and the very deep, long-term dimming events in the light curves. Since the dimming events show little to no structure (see Figure \ref{figure:phased}), it is likely the disk around the hot companion is not only large but almost completely opaque. Also, if the secondary component is an sdB star with a large disk, and the M-giant is roughly three orders of magnitude brighter in optical flux than the sdB star (see Figure \ref{SED_Fit}), the secondary eclipse would be $\sim$1 mmag in depth, and thus undetectable in any of our data sets. 

To determine some of the physical properties of the opaque eclipsing body, we model the 2011--2014 eclipse as an occultation of the M-giant by a large opaque object with a sharp leading knife-edge, perpendicular to its direction of motion. This model requires no knowledge of the orbital eccentricity. Using this simple model, we can calculate a transverse velocity of the occulting body 2$\times$R$_{Star}$/T, where T is the estimated ingress or egress timescale. \citet{Afanasiev:2013} measured the spectra of the primary star to be consistent with an M-giant. The stellar radius of an M-giant ranges from $\sim$45 R$_{\sun}$ (M0 III) to $\sim$170 R$_{\sun}$ (M7/8 III) \citep{Dumm:1998}. We estimate the egress of the 2011-2014 eclipse to be $\sim$315 days. This translates to a range of transverse velocity of 2.3--8.7 \kms (for the range of stellar radii). Using the total estimated duration of the eclipse to be $\sim$3.45 years, we also estimate the extent of the occulting body to be V$\times$T(duration) = 1.7--6.3 AU (the disk could be inclined with respect to the companion's orbital motion resulting in a larger disk). Combining the estimated period of the EB (69.068 years) with a mass estimate for the M-giant and the hot companion, we can estimate the semi-major axis of the system, assuming Keplerian motion and a circular orbit. For the hot companion, we adopt a white dwarf mass range of 0.17--1.33 M$_{\sun}$ \citep{Kepler:2007, Kilic:2007} and M-giants can range from 0.8 to 5.0 M$_{\sun}$ \citep{Bressan:1993, Dumm:1998}. Using these mass ranges, this would result in a semi-major axis range of 16.7--31.2 AU. By applying a simple model, we are able to determine that the occulting body is moving 2.3--8.7 \kms, is 1.7--6.3 AU wide, and is orbiting at a semi-major axis of 16.7--31.2 AU. This would suggest that the hot companion has a few AU diameter disk around it. The 4.5 mag depth of the eclipse implies that the occulter almost completely occults the M-giant. Therefore, if the disk in not inclined to our line-of-sight, the thickness of the disk must be similar to the diameter of the M-giant (45--170 R$_{\sun}$ or 0.21--0.8 AU). It is possible that the disk is inclined to our line-of-sight ($\sim89^{\circ}\pm1.0^{\circ}$ for $\epsilon$ Aur, see \citet{Kloppenborg:2015}). If the disk is not edge-on, the thickness of the disk could be significantly thinner (or even thicker) and still cause the eclipse seen. Therefore, we are not able to constrain the disk's thickness. In the case of an edge on disk, the disk thickness-to-diameter ratio would be $\sim$12\%.


\subsection{Alternate Explanations}
We have presented evidence in the previous subsection that the large dimming events of TYC 2505-672-1 are caused by the M-giant primary being eclipsed by a white dwarf with a large disk surrounding it. We now explore an alternate explanation for these observations.

Another possible explanation for such large dimming events is that the M-giant primary is an RCB star or entering the RCB phase \citep{Denisenko:2013}. These are carbon rich supergiants (usually F or G spectral types) that experience non-periodic, large dimming events (up to $\sim$8 mag in depth) caused by the formation of carbon dust in the stellar atmosphere. The dimmings are typically separated by a few years to a decade and are typically $>$3 mag in depth. The drop in the RCB star's brightness is very rapid (a few days to weeks) while the recovery is much slower (months to years). These stars also are known to pulsate with amplitudes of $\sim$0.1 mag \citep{Clayton:2012}. If we were to believe that TYC 2505-672-1 was a unknown RCB star, the UV excess seen in the SED would be from a faint white dwarf orbiting it contributing some UV flux. Since the dimmings are separated by much longer then a few years to a decade, there is no pulsation amplitude observed outside of the most recent dimming (where we have the best photometric precision), the spectra observed by \citet{Denisenko:2013} indicate that the primary star is an M-giant (not a supergiant, and the SED analysis supports this), and the most recent dimming show the ingress/egress timescales to be both much longer and more uniform than observed in known RCB stars, we do not believe the RCB scenario to be a plausible explanation for the dimming events observed for TYC 2505-672-1.

\section{Summary and Conclusions}
We have presented new observations of the remarkable eclipsing system TYC 2505-672-1, an M-giant star that has shown two separate dimming events separated by $\sim$69.1 yr over the course of the historical light curve spanning 120 yr. We find that both eclipses phase up nicely with a period of 69.068 years. The most recent event, which was observed by KELT-North, CRTS, AAVSO, and ASAS-SN, show that the eclipse lasts $\sim$3.5 years, has a depth of 4.5 mag in the optical, and shows little to no structure in the lightcurve during the eclipse. Our SED analysis (both in and out of eclipse) indicates to two components, one with a T$_{\rm eff}$=3600 K and the other with T$_{\rm eff}$=8000 K. Combining the SED and photometric analysis, we determine that the system contains an M-giant primary star and a hot, dim companion. 

Curiously, however, the hot companion has a radius that is much too small to be a main-sequence dwarf and much too large to be a standard cooling white dwarf. We propose the best solution is that the M-giant is being eclipsed every $\sim$69.1 years by a ``stripped red giant" (pre-Helium-white-dwarf, low-mass subdwarf B-type) companion surrounded by a large, opaque disk. This would explain the UV excess in the SED and the near-total occultation seen in the photometry, while also explaining the seemingly strange radius of the hot component. 

As with $\epsilon$ Aurigae, this system presents a unique laboratory for understanding the disk structure of a companion orbiting an evolved star. At a orbital period of $\sim$69.1 years, this is now the longest period eclipsing system found to date. We encourage continued photometric and spectroscopic follow-up of this system, in particular the measurement of the system's radial velocity motion. Extrapolating from our calculated period and T$_C$, the next eclipse should begin in early UT 2080 April and end in mid UT 2083 September (T$_C$ = 2480857.48, UT 2081 December 24).

A mystery remains regarding the evolutionary nature of the hot component within the opaque disk. Previous examples of pre-He-WDs \citep[e.g.,][]{Maxted:2014} are in relatively short-period binary systems (periods of $\sim$1 day), such that the recent stripping of the red giant that produced the currently observed hot source can be reasonably attributed to interactions between the close binary components. In the present case, however, the two stars are evidently very widely separated (semi-major axis $\sim$20 AU). Perhaps the hot component is itself in a tight binary within the surrounding opaque disk, or is the result of a white dwarf merger. It is possible that we are witnessing an object in the very short-lived evolutionary state following the sdB stage leading to the eventual very hot, and then cooling, white dwarf.

\acknowledgments
Early work on KELT-North was supported by NASA Grant NNG04GO70G. J.A.P. and K.G.S. acknowledge support from the Vanderbilt Office of the Provost through the Vanderbilt Initiative in Data-intensive Astrophysics. This work has made use of NASA's Astrophysics Data System and the SIMBAD database operated at CDS, Strasbourg, France.

Work by B.S.G. and D.J.S. was partially supported by NSF CAREER Grant AST-1056524. Work by K.G.S. was supported by NSF PAARE grant AST-1358862. 

The DASCH project at Harvard is grateful for partial support from NSF grants AST-0407380, AST-0909073, and AST-1313370.

The CSS survey is funded by the National Aeronautics and Space Administration under Grant No. NNG05GF22G issued through the Science Mission Directorate Near-Earth Objects Observations Program. The CRTS survey is supported by the U.S.~National Science Foundation under grants AST-0909182 and AST-1313422.

Development of ASAS-SN has been supported by NSF grant AST-0908816 and CCAPP at the Ohio State University.  ASAS-SN is supported by NSF grant AST-1515927, the Center for Cosmology and AstroParticle Physics (CCAPP) at OSU, the Mt. Cuba Astronomical Foundation, George Skestos, and the Robert Martin Ayers Sciences Fund.

B.S. is supported by NASA through Hubble Fellowship grant HF-51348.001 awarded by the Space Telescope Science Institute, which is operated by the Association of Universities for Research in Astronomy, Inc., for NASA, under contract NAS 5-26555.  CSK is supported by NSF grants AST-1515876 and AST-1515927.

\bibliographystyle{apj}

\bibliography{Eps2}

\begin{thebibliography}{}
\expandafter\ifx\csname natexlab\endcsname\relax\def\natexlab#1{#1}\fi

\bibitem[{{Afanasiev} {et~al.}(2013){Afanasiev}, {Denisenko}, {Krushinsky},
  {Gorbovskoy}, {Lipunov}, {Balanutsa}, {Yecheistov}, {Tiurina}, {Kornilov},
  {Belinski}, {Shatskiy}, {Chazov}, {Kuznetsov}, {Zimnukhov}, {Zalozhnih},
  {Popov}, {Bourdanov}, {Punanova}, {Ivanov}, {Yazev}, {Budnev},
  {Konstantinov}, {Chuvalaev}, {Poleshchuk}, {Gress}, {Parkhomenko}, {Tlatov},
  {Dormidontov}, {Senik}, {Yurkov}, {Sergienko}, {Varda}, {Sinyakov},
  {Shurpakov}, {Shumkov}, {Podvorotny}, {Levato}, {Saffe}, {Mallamaci},
  {Lopez}, \& {Podest}}]{Afanasiev:2013}
{Afanasiev}, V., {Denisenko}, D., {Krushinsky}, V., {et~al.} 2013, The
  Astronomer's Telegram, 4834, 1

\bibitem[{{Bertin} \& {Arnouts}(1996)}]{Bertin:1996}
{Bertin}, E., \& {Arnouts}, S. 1996, \aaps, 117, 393

\bibitem[{{Bianchi} {et~al.}(2011){Bianchi}, {Herald}, {Efremova}, {Girardi},
  {Zabot}, {Marigo}, {Conti}, \& {Shiao}}]{Bianchi:2011}
{Bianchi}, L., {Herald}, J., {Efremova}, B., {et~al.} 2011, \apss, 335, 161

\bibitem[{{Bressan} {et~al.}(1993){Bressan}, {Fagotto}, {Bertelli}, \&
  {Chiosi}}]{Bressan:1993}
{Bressan}, A., {Fagotto}, F., {Bertelli}, G., \& {Chiosi}, C. 1993, \aaps, 100,
  647

\bibitem[{{Carroll} {et~al.}(1991){Carroll}, {Guinan}, {McCook}, \&
  {Donahue}}]{Carroll:1991}
{Carroll}, S.~M., {Guinan}, E.~F., {McCook}, G.~P., \& {Donahue}, R.~A. 1991,
  \apj, 367, 278

\bibitem[{{Clayton}(2012)}]{Clayton:2012}
{Clayton}, G.~C. 2012, Journal of the American Association of Variable Star
  Observers (JAAVSO), 40, 539

\bibitem[{{Cutri} \& {et al.}(2014)}]{Cutri:2014}
{Cutri}, R.~M., \& {et al.} 2014, VizieR Online Data Catalog, 2328, 0

\bibitem[{{Cutri} {et~al.}(2003){Cutri}, {Skrutskie}, {van Dyk}, {Beichman},
  {Carpenter}, {Chester}, {Cambresy}, {Evans}, {Fowler}, {Gizis}, {Howard},
  {Huchra}, {Jarrett}, {Kopan}, {Kirkpatrick}, {Light}, {Marsh}, {McCallon},
  {Schneider}, {Stiening}, {Sykes}, {Weinberg}, {Wheaton}, {Wheelock}, \&
  {Zacarias}}]{Cutri:2003}
{Cutri}, R.~M., {Skrutskie}, M.~F., {van Dyk}, S., {et~al.} 2003, VizieR Online
  Data Catalog, 2246, 0

\bibitem[{{Denisenko} {et~al.}(2013){Denisenko}, {Gorbovskoy}, {Lipunov},
  {Balanutsa}, {Yecheistov}, {Tiurina}, {Kornilov}, {Belinski}, {Shatskiy},
  {Chazov}, {Kuznetsov}, {Zimnukhov}, {Krushinsky}, {Zalozhnih}, {Popov},
  {Bourdanov}, {Punanova}, {Ivanov}, {Yazev}, {Budnev}, {Konstantinov},
  {Chuvalaev}, {Poleshchuk}, {Gress}, {Parkhomenko}, {Tlatov}, {Dormidontov},
  {Senik}, {Yurkov}, {Sergienko}, {Varda}, {Sinyakov}, {Shurpakov}, {Shumkov},
  {Podvorotny}, {Levato}, {Saffe}, {Mallamaci}, {Lopez}, \&
  {Podest}}]{Denisenko:2013}
{Denisenko}, D., {Gorbovskoy}, E., {Lipunov}, V., {et~al.} 2013, The
  Astronomer's Telegram, 4784, 1

\bibitem[{{Drake} {et~al.}(2009){Drake}, {Djorgovski}, {Mahabal}, {Beshore},
  {Larson}, {Graham}, {Williams}, {Christensen}, {Catelan}, {Boattini},
  {Gibbs}, {Hill}, \& {Kowalski}}]{Drake:2009}
{Drake}, A.~J., {Djorgovski}, S.~G., {Mahabal}, A., {et~al.} 2009, \apj, 696,
  870

\bibitem[{{Dumm} \& {Schild}(1998)}]{Dumm:1998}
{Dumm}, T., \& {Schild}, H. 1998, NA, 3, 137

\bibitem[{{Grindlay} {et~al.}(2012){Grindlay}, {Tang}, {Los}, \&
  {Servillat}}]{Grindley:2012}
{Grindlay}, J., {Tang}, S., {Los}, E., \& {Servillat}, M. 2012, in IAU
  Symposium, Vol. 285, IAU Symposium, ed. E.~{Griffin}, R.~{Hanisch}, \&
  R.~{Seaman}, 29--34

\bibitem[{{Henden} {et~al.}(2015){Henden}, {Levine}, {Terrell}, \&
  {Welch}}]{Henden:2015}
{Henden}, A.~A., {Levine}, S., {Terrell}, D., \& {Welch}, D.~L. 2015, in
  American Astronomical Society Meeting Abstracts, Vol. 225, American
  Astronomical Society Meeting Abstracts, 336.16

\bibitem[{{Hoard} {et~al.}(2010){Hoard}, {Howell}, \& {Stencel}}]{Hoard:2010}
{Hoard}, D.~W., {Howell}, S.~B., \& {Stencel}, R.~E. 2010, \apj, 714, 549

\bibitem[{{Hog} {et~al.}(1998){Hog}, {Kuzmin}, {Bastian}, {Fabricius},
  {Kuimov}, {Lindegren}, {Makarov}, \& {Roeser}}]{Hog:1998}
{Hog}, E., {Kuzmin}, A., {Bastian}, U., {et~al.} 1998, \aap, 335, L65

\bibitem[{{H{\o}g} {et~al.}(2000){H{\o}g}, {Fabricius}, {Makarov}, {Urban},
  {Corbin}, {Wycoff}, {Bastian}, {Schwekendiek}, \& {Wicenec}}]{Hog:2000}
{H{\o}g}, E., {Fabricius}, C., {Makarov}, V.~V., {et~al.} 2000, \aap, 355, L27

\bibitem[{{Kepler} {et~al.}(2007){Kepler}, {Kleinman}, {Nitta}, {Koester},
  {Castanheira}, {Giovannini}, {Costa}, \& {Althaus}}]{Kepler:2007}
{Kepler}, S.~O., {Kleinman}, S.~J., {Nitta}, A., {et~al.} 2007, \mnras, 375,
  1315

\bibitem[{{Kilic} {et~al.}(2007){Kilic}, {Allende Prieto}, {Brown}, \&
  {Koester}}]{Kilic:2007}
{Kilic}, M., {Allende Prieto}, C., {Brown}, W.~R., \& {Koester}, D. 2007, \apj,
  660, 1451

\bibitem[{{Kloppenborg} {et~al.}(2010){Kloppenborg}, {Stencel}, {Monnier},
  {Schaefer}, {Zhao}, {Baron}, {McAlister}, {ten Brummelaar}, {Che},
  {Farrington}, {Pedretti}, {Sallave-Goldfinger}, {Sturmann}, {Sturmann},
  {Thureau}, {Turner}, \& {Carroll}}]{Kloppenborg:2010}
{Kloppenborg}, B., {Stencel}, R., {Monnier}, J.~D., {et~al.} 2010, \nat, 464,
  870

\bibitem[{{Kloppenborg} {et~al.}(2015){Kloppenborg}, {Stencel}, {Monnier},
  {Schaefer}, {Baron}, {Tycner}, {Zavala}, {Hutter}, {Zhao}, {Che}, {ten
  Brummelaar}, {Farrington}, {Parks}, {McAlister}, {Sturmann}, {Sturmann},
  {Sallave-Goldfinger}, {Turner}, {Pedretti}, \& {Thureau}}]{Kloppenborg:2015}
{Kloppenborg}, B.~K., {Stencel}, R.~E., {Monnier}, J.~D., {et~al.} 2015, \apjs,
  220, 14

\bibitem[{{Lipunov} {et~al.}(2010){Lipunov}, {Kornilov}, {Gorbovskoy},
  {Shatskij}, {Kuvshinov}, {Tyurina}, {Belinski}, {Krylov}, {Balanutsa},
  {Chazov}, {Kuznetsov}, {Kortunov}, {Sankovich}, {Tlatov}, {Parkhomenko},
  {Krushinsky}, {Zalozhnyh}, {Popov}, {Kopytova}, {Ivanov}, {Yazev}, \&
  {Yurkov}}]{Lipunov:2010}
{Lipunov}, V., {Kornilov}, V., {Gorbovskoy}, E., {et~al.} 2010, Advances in
  Astronomy, 2010, 30

\bibitem[{{Maxted} {et~al.}(2014){Maxted}, {Serenelli}, {Marsh}, {Catal{\'a}n},
  {Mahtani}, \& {Dhillon}}]{Maxted:2014}
{Maxted}, P.~F.~L., {Serenelli}, A.~M., {Marsh}, T.~R., {et~al.} 2014, \mnras,
  444, 208

\bibitem[{{Pepper} {et~al.}(2007){Pepper}, {Pogge}, {DePoy}, {Marshall},
  {Stanek}, {Stutz}, {Poindexter}, {Siverd}, {O'Brien}, {Trueblood}, \&
  {Trueblood}}]{Pepper:2007}
{Pepper}, J., {Pogge}, R.~W., {DePoy}, D.~L., {et~al.} 2007, \pasp, 119, 923

\bibitem[{{Pickles} \& {Depagne}(2010)}]{Pickles:2010}
{Pickles}, A., \& {Depagne}, {\'E}. 2010, \pasp, 122, 1437

\bibitem[{{Robitaille} {et~al.}(2006){Robitaille}, {Whitney}, {Indebetouw},
  {Wood}, \& {Denzmore}}]{Robitaille:2006}
{Robitaille}, T.~P., {Whitney}, B.~A., {Indebetouw}, R., {Wood}, K., \&
  {Denzmore}, P. 2006, \apjs, 167, 256

\bibitem[{{Schlegel} {et~al.}(1998){Schlegel}, {Finkbeiner}, \&
  {Davis}}]{Schlegel:1998}
{Schlegel}, D.~J., {Finkbeiner}, D.~P., \& {Davis}, M. 1998, \apj, 500, 525

\bibitem[{{Shappee} {et~al.}(2014){Shappee}, {Prieto}, {Grupe}, {Kochanek},
  {Stanek}, {De Rosa}, {Mathur}, {Zu}, {Peterson}, {Pogge}, {Komossa}, {Im},
  {Jencson}, {Holoien}, {Basu}, {Beacom}, {Szczygie{\l}}, {Brimacombe},
  {Adams}, {Campillay}, {Choi}, {Contreras}, {Dietrich}, {Dubberley},
  {Elphick}, {Foale}, {Giustini}, {Gonzalez}, {Hawkins}, {Howell}, {Hsiao},
  {Koss}, {Leighly}, {Morrell}, {Mudd}, {Mullins}, {Nugent}, {Parrent},
  {Phillips}, {Pojmanski}, {Rosing}, {Ross}, {Sand}, {Terndrup}, {Valenti},
  {Walker}, \& {Yoon}}]{Shappee:2014}
{Shappee}, B.~J., {Prieto}, J.~L., {Grupe}, D., {et~al.} 2014, \apj, 788, 48

\bibitem[{{Siverd} {et~al.}(2012){Siverd}, {Beatty}, {Pepper}, {Eastman},
  {Collins}, {Bieryla}, {Latham}, {Buchhave}, {Jensen}, {Crepp}, {Street},
  {Stassun}, {Gaudi}, {Berlind}, {Calkins}, {DePoy}, {Esquerdo}, {Fulton},
  {F{\H u}r{\'e}sz}, {Geary}, {Gould}, {Hebb}, {Kielkopf}, {Marshall}, {Pogge},
  {Stanek}, {Stefanik}, {Szentgyorgyi}, {Trueblood}, {Trueblood}, {Stutz}, \&
  {van Saders}}]{Siverd:2012}
{Siverd}, R.~J., {Beatty}, T.~G., {Pepper}, J., {et~al.} 2012, \apj, 761, 123

\bibitem[{{Tang} {et~al.}(2013){Tang}, {Grindlay}, {Bildsten}, \&
  {collaborators}}]{Tang:2013}
{Tang}, S., {Grindlay}, J.~E., {Bildsten}, L., \& {collaborators}, m. 2013, in
  Giants of Eclipse, 20302

\end{thebibliography}

\end{document}